\documentclass[12pt]{article}
\usepackage{makeidx}
\usepackage{amsmath}
\usepackage{amsfonts}
\usepackage{amssymb}
\usepackage{epsfig}
\usepackage[cp1251]{inputenc}
\usepackage{graphicx}

\newcounter{defin}
\newcounter{lemma}
\newcounter{theorem}
\newcounter{proposition}
\newcounter{example}

\begin{document}

\title{Log-Sobolev inequality and proof of Hypothesis of the Gaussian
Maximizers for the capacity of quantum noisy homodyning}
\author{A. S. Holevo \\
%EndAName
Steklov Mathematical Institute, RAS, Moscow, Russia}
\date{}
\maketitle

\begin{abstract}
In the present paper we give proof that the information-transmission
capacity of the approximate position measurement with the oscillator energy
constraint, which underlies noisy Gaussian homodyning in quantum optics, is
attained on Gaussian encoding. The proof is based on general principles of
convex programming. Rather remarkably, for this particular model the method
reduces the solution of the optimization problem to a generalization of the
celebrated log-Sobolev inequality. We hope that this method should work also
for other models lying out of the scope of the ``threshold condition''
ensuring that the upper bound for the capacity as a difference between the
maximum and the minimum output entropies is attainable.
\end{abstract}

\section{Introduction}

Quantum Shannon theory was rapidly developing during the past decades. As
distinct from the classical case, quantum channel is characterized by a
whole variety of different capacities, depending on the type of transmitted
information (classical or quantum) and on additional resources used during
transmission. Among these capacities, the \textit{classical capacity}, i.e.
the ultimate rate of reliable transmission of classical data via a quantum
channel, plays a distinguished role, both historically and because of its
central importance for quantum communications. A long-standing problem is
the classical capacity of bosonic Gaussian channels of various kinds.
Hypothesis of Gaussian Maximizers (HGM) states that the full capacity of
such channels is attained on Gaussian encodings. The conjecture turned out
remarkably difficult. After a series of intermediate steps, a breakthrough
was made in the papers \cite{ghg}, \cite{ghm}, \cite{ghm1}, where HGM was
proved for important class of multimode gauge co- or contra-variant channels
(called phase-insensitive in quantum optics). In \cite{h2}, \cite{acc-noJ}
the solution was extended to a much broader class of channels satisfying
certain ``threshold condition'', essentially ensuring that the upper bound
for the capacity as a difference between the maximum and the minimum output
entropies is attainable. In \cite{h2}, \cite{acc2} these findings were
extended to Gaussian measurement (quantum-classical) channels. At the same
time, HGM remains open for rather large variety of quantum and
quantum-classical channels lying beyond the scope of the threshold condition
\cite{entropy}. At this point we want to mention that many authors (see e.g.
\cite{hall3}, \cite{scha}, \cite{guha}, \cite{lee}) explored the maximum
information restricting to the class of Gaussian encodings. In the absence
of proof of HGM such results give only a lower bound for the full classical
capacity of the Gaussian channel.

Among these models there are rather elementary Gaussian channels, such as
approximate position measurement with the oscillator energy constraint,
which essentially underlies noisy Gaussian homodyning in quantum optics.
While this and related models were studied in earlier quantum communication
papers (see e.g. \cite{caves}, \cite{hall2}), we could not find a proof of
HGM for them in the literature. In the present paper we give such a proof
which is based on general principles of convex programming. Rather
remarkably, for this particular model the method reduces the solution of the
optimization problem to a generalization of log-Sobolev inequality \cite%
{ledoux}, \cite{lieb}-- one of the highlights of Analysis in the past
century. We think that this method should work also for other models out of
the scope of the threshold condition, and hope to address it in further
publications.

\section{The classical capacity of quantum measurement}

\label{s1}

Let $\mathcal{H}$ be a Hilbert space of quantum system. Statistics of a
quantum measurement with the outcomes $y\in\mathcal{Y}$ is described by a
probability operator-valued measure $M=\{M(dy)\}$ (POVM) on $\mathcal{H}$,
that is $M(dy)\geq 0$ and $\int M(dy)=I$ (the unit operator on $\mathcal{H}$%
). As shown e.g. in \cite{acc}, under mild separability assumptions there
exists a countably finite measure $\mu (dy)$ such that for any density
operator $\rho $ the distribution of the measurement outcomes $\mathrm{Tr}%
\rho M(dy)$ is absolutely continuous w.r.t. $\mu (dy),$ thus having the
probability density $p_{\rho }(y)$. The affine map $M:\rho \rightarrow
p_{\rho }(\cdot )$ will be called the \textit{measurement channel}.

An \textit{ensemble} (encoding) $\mathcal{E}=\left\{ \pi (dx),\rho
(x)\right\} $ consists of probability measure $\pi (dx)$ on a ``alphabet'' $%
\mathcal{X}$ and a measurable family of density operators (quantum states) $%
x\rightarrow \rho (x)$ on the Hilbert space $\mathcal{H}$ of the quantum
system. The \textit{average state} of the ensemble is the barycenter of this
measure%
\begin{equation*}
\bar{\rho}_{\mathcal{E}}=\int_{\mathcal{X}}\rho (x)\,\pi (dx).
\end{equation*}%
%
%
%the integral existing in the strong sense in the Banach space of trace-class
%operators on $\mathcal{H}$.

The classical Shannon information between the input $x$ and the measurement
outcome $y$ is equal to%
\begin{equation*}
I(\mathcal{E},M)=\int \int \pi (dx)\mu (dy)p_{\rho (x)}(y)\ln \frac{p_{\rho
(x)}(y)}{p_{\bar{\rho}_{\mathcal{E}}}(y)}
\end{equation*}%
In what follows we will consider POVMs having (uniformly) bounded operator
density, $M(dy)=m(y)\mu (dy),$ with $\left\Vert m(y)\right\Vert \leq b,$ so
that the probability densities $p_{\rho }(y)=\mathrm{Tr}\,\rho m(y)$ are
uniformly bounded, $0\leq p_{\rho }(y)\leq b$. %(The probability densities
%corresponding to Gaussian observables we will be dealing with possess this property).
Moreover, by including $b$ into $\mu (dy),$ we can assume without loss of
generality that $b=1.$ Then the output differential entropy
\begin{equation}
h_{M}(\rho )=-\int p_{\rho }(y)\ln \,p_{\rho }(y)\mu (dy)  \label{den}
\end{equation}%
is well defined with values in $[0,+\infty ]$ (see \cite{acc} for the
detail). $.$

Next we define the quantity (\ref{den}):
\begin{equation}
e_{M}(\rho )=\inf_{\mathcal{E}:\bar{\rho}_{\mathcal{E}}=\rho }\int
h_{M}(\rho (x))\pi (dx),  \label{e1}
\end{equation}%
which is an analog of the \textit{convex closure of the output differential
entropy} for a quantum channel \cite{Shir}.

Let $H$ be a Hamiltonian in the Hilbert space $\mathcal{H}$ of the quantum
system, $E$ a positive number. Then the \textit{energy-constrained classical
capacity} of the measurement channel $M$ is
\begin{equation}
C(M,H,E)=\sup_{\mathcal{E}:\mathrm{Tr}\bar{\rho}_{\mathcal{E}}H\leq E}I(%
\mathcal{E},M),  \label{0}
\end{equation}%
where maximization is over the input ensembles of states $\mathcal{E}$
satisfying the energy constraint $\mathrm{Tr}\bar{\rho}_{\mathcal{E}}H\leq E$%
. As shown in \cite{acc-noJ}, proposition 1, it is indeed the capacity in
the sense of information theory, i.e. the ultimate rate of the classical
asymptotically reliable data transmission via the measurement channel. Note
that the measurement channel is \textit{entanglement-breaking} \cite{QSCI}
hence its classical capacity is additive and is given by the one-shot
expression (\ref{0}).

If $h_{M}(\bar{\rho}_{\mathcal{E}})<+\infty $, then
\begin{equation}
I(\mathcal{E},M)=h_{M}(\bar{\rho}_{\mathcal{E}})-\int h_{M}(\rho (x))\pi
(dx).  \label{i1}
\end{equation}%
By using (\ref{i1}), (\ref{e1}), we obtain%
\begin{eqnarray}
C(M,H,E) &=&\sup_{\rho :\mathrm{Tr}\rho H\leq E}\left[ h_{\mathcal{M}}(\rho
)-e_{M}(\rho )\right]  \label{cmfe} \\
&\leq &\sup_{\rho :\mathrm{Tr}\rho H\leq E}h_{M}(\rho )-\inf_{\rho
}h_{M}(\rho ).  \notag
\end{eqnarray}%
There are important cases where the last inequality turns into equality thus
giving the value of the capacity. This happens when the maximizer of the
first term can be represented as a mixture of (pure) states minimizing $%
h_{M}(\rho )$ \cite{h5}, \cite{acc-noJ}. In particular, all the instances
where the Hypothesis of Gaussian Maximizer was proved for Gaussian channels
so far refer to that case. In the present paper we propose a method allowing
to prove this hypothesis for the first case where this condition is violated
and the inequality in (\ref{cmfe}) is strict and hence becomes useless. In
sec. \ref{s4} we illustrate it on the example of approximate position
measurement (corresponding to noisy homodyning in quantum optics) where this
violation happens in the most extreme form.

\section{Reduction to a convex programming problem}

\label{s2}

In this section we propose a method of computation of the quantity $%
e_{M}(\rho )$ based on similarity of the optimization problem in the right
side of (\ref{e1}) and general quantum Bayes estimation problem \cite{jap},
\cite{jma1}, \cite{hol}. Consider a measurement channel given by the map $%
M:\rho \rightarrow p_{\rho }(y)=\mathrm{Tr\,}\rho m(y),$ where $m(y)$ is a
uniformly bounded positive-operator-valued function of $y\in \mathcal{Y},$
such that $\int m(y)\mu (dy)=I$. Any ensemble $\mathcal{E}=\left\{ \pi
(dx),\rho (x)\right\} $ where $x\in \mathcal{X},$ can be equivalently
considered as a probability distribution $\pi (d\rho )$ on the whole set of
quantum states $\mathfrak{S=S}(\mathcal{H})$ (with the carrier concentrated
of the states $\rho (x),x\in \mathcal{X}$). Another equivalent description
of $\mathcal{E}$ is given by the positive (but not probability!)
operator-valued measure $\Pi (d\rho )=\rho \pi (d\rho )$ with values in $%
\mathfrak{S.}$ The average state is then%
\begin{equation*}
\bar{\rho}_{\mathcal{E}}=\int_{\mathfrak{S}}\rho \pi (d\rho )=\Pi (\mathfrak{%
S}),
\end{equation*}%
and the minimized functional%
\begin{equation*}
F(\mathcal{E})=\int_{\mathfrak{S}}h\left( p_{\rho }\right) \pi (d\rho
)=\int_{\mathfrak{S}}\mathrm{Tr\,}\,K(\rho )\Pi (d\rho ),
\end{equation*}%
where%
\begin{equation}
K(\rho )=-\int m(y)\ln p_{\rho }(y)\mu (dy).  \label{K}
\end{equation}%
By fixing an average state $\overline{\rho },$ we arrive at the optimization
problem%
\begin{eqnarray*}
\int_{\mathfrak{S}}\mathrm{Tr\,}K(\rho )\Pi (d\rho ) &\longrightarrow &\min
\\
\Pi (d\rho) &\geq &0 \\
\Pi (\mathfrak{S}) &=&\overline{\rho },
\end{eqnarray*}%
which is formally similar to one arising in the general quantum Bayes
problem \cite{jma1}, \cite{hol}. The minimized functional is affine in $%
\mathcal{E}=\{\Pi (d\rho)\}$ and the constraints are convex, so it is a
convex programming problem. Under certain regularity condition the problem
was investigated in \cite{hol}, where the following necessary and sufficient
conditions for optimality of an ensemble $\mathcal{E} _{0}=\{\Pi _{0}(d\rho
)\}$ were given, which we reproduce here \textit{formally}:

\textit{There exists Hermitian operator }$\Lambda _{0}$\textit{\ such that}

(i) $\Lambda _{0}\leq K(\rho )$ \textit{for all} $\rho \in \mathfrak{S};$

(ii) $\left[ K(\rho )-\Lambda _{0}\right] \,\Pi _{0}(d\rho )=0$.

Moreover, $\Lambda _{0}$ is the solution of the dual problem%
\begin{equation}
\max \left\{ \mathrm{Tr\,}\overline{\rho }\Lambda :\Lambda ^{\ast }=\Lambda
,\,\Lambda \leq K(\rho )\text{ for all }\rho \in \mathfrak{S}\right\} .
\end{equation}

By integrating (ii), we obtain the equation for determination of $\Lambda
_{0}$
\begin{equation}
\int_{\mathfrak{S}}K(\rho )\,\Pi _{0}(d\rho )=\int_{\mathfrak{S}}K(\rho
)\,\rho \,\pi _{0}(d\rho )=\Lambda _{0}\overline{\rho }.  \label{iii}
\end{equation}%
The sufficiency of the conditions (i), (ii) is easy to demonstrate formally
(cf. \cite{jap}): for any $\mathcal{E}=\left\{ \Pi (d\rho )\right\} $
\begin{eqnarray*}
F(\mathcal{E}) &=&\int_{\mathfrak{S}}\mathrm{Tr\,}K(\rho )\Pi (d\rho )%
\overset{(i)}{\geq }\int_{\mathfrak{S}}\mathrm{Tr\,}\Lambda _{0}\Pi (d\rho )
\\
&=&\mathrm{Tr\,}\Lambda _{0}\overline{\rho }=\int_{\mathfrak{S}}\mathrm{Tr\,}%
\Lambda _{0}\Pi _{0}(d\rho )\overset{(ii)}{=}\int_{\mathfrak{S}}\mathrm{Tr\,}%
K(\rho )\Pi _{0}(d\rho )=F(\mathcal{E}_{0}).
\end{eqnarray*}

Coming back to the parametric representation $\mathcal{E}=\left\{ \pi
(dx),\rho (x)\right\} ,$ we can write the condition (ii) as%
\begin{equation*}
\left[ K(\rho (x))-\Lambda _{0}\right] \,\rho (x)=0,\quad \mathrm{a.e.}~x\in
\mathcal{X},
\end{equation*}%
which means that the equality holds a.e. with respect to the measure $\pi
_{0}(dx).$ The equation (\ref{iii}) becomes%
\begin{equation}  \label{KL}
\int K(\rho (x))\,\rho (x)\,\pi _{0}(dx)=\Lambda _{0}\overline{\rho }.
\end{equation}%
In the case of the measurement channel $M:\rho \rightarrow p_{\rho }(y)=%
\mathrm{Tr\,}\rho m(y)$, this equation reduces to%
\begin{equation*}
-\int \int m(y)\ln p_{\rho (x)}(y)\,\rho (x)\,\mu (dy)\,\pi _{0}(dx)=\Lambda
_{0}\overline{\rho }.
\end{equation*}

In specific applications, like HGM for bosonic Gaussian channel, the
candidate for an optimal encoding usually can be found by optimizing in the
class of Gaussian encodings. Then the condition (ii) for this candidate can
be verified and the operator $\Lambda_0$ found, while a major difficulty may
be the check of the operator inequality (i).

\section{Gaussian systems}

\label{s3}

We will systematically use some notations and results from the books \cite%
{asp}, \cite{QSCI}. Consider the finite-dimensional symplectic vector space $%
(Z,\Delta )$ with $Z=\mathbb{R}^{2s}$ and the canonical symplectic matrix
\begin{equation}
\Delta =\mathrm{diag}\left[
\begin{array}{cc}
0 & 1 \\
-1 & 0%
\end{array}%
\right] _{j=1,\dots ,s}.
\end{equation}%
In what follows $\mathcal{H}$ will be the space of an irreducible
representation $z\rightarrow W(z);\,z\in Z,$ of the \textit{canonical
commutation relations (CCR)} %\begin{equation}
\begin{equation}
W(z)W(z^{\prime })=\exp [-\frac{i}{2}z^{t}\Delta z^{\prime }]\,W(z+z^{\prime
}).  \label{eq17}
\end{equation}%
%
%
%\end{equation}
Here $W(z)=\exp i\,Rz$ are the unitary Weyl operators with the generators
\begin{equation}
Rz=\sum_{j=1}^{s}(x_{j}q_{j}+y_{j}p_{j}),\quad z=[x_{j}\,\,y_{j}]_{j=1,\dots
,s}^{t}
\end{equation}%
and $R=\left[ q_{1},p_{1},\dots ,q_{s},p_{s}\right] $ is the row vector of
the bosonic position-momentum observables, satisfying the canonical
commutation relation
\begin{equation*}
q_{j}p_{k}-p_{k}q_{j}\subseteq i\delta _{jk}I,\quad j,k=1,\dots ,s.\quad
\end{equation*}%
In quantum communication theory $q_{j},p_{j}$ describe the relevant modes of
the field on receiver's aperture (see, e.g. \cite{sera}). A number of
analytical complications related to unboundedness of operators unavoidably
arises in connection with Bosonic systems and CCR. In our treatment of CCR
we focus on the algebraic aspects essential for applications while a
presentation of the related analytical detail such as domains of definition,
selfadjointness etc. can be found in the literature\footnote{%
See e.g. \cite{asp}, \cite{QSCI} for the detail of mathematical treatment of
expressions with the unbounded operators related to $R.$}

The \textit{displacement operators} $D(z)=W(-\Delta ^{-1}z)$ satisfy the
equation that follows from the canonical commutation relations (\ref{eq17})
\begin{equation}
D(z)^{\ast }W(w)D(z)=\exp \left( iw^{t}z\right) W(w).  \label{dis}
\end{equation}%
The quantum Fourier transform of a trace class operator $\rho $ is defined
as
\begin{equation*}
\mathop{\rm Tr}\nolimits\rho D(w)
\end{equation*}%
When $\rho $ is a density operator, this is called the quantum
characteristic function of the state $\rho .$ The quantum Parceval formula
holds \cite{asp}:%
\begin{equation}
\mathop{\rm Tr}\nolimits\rho \sigma ^{\ast }=\int \mathop{\rm Tr}%
\nolimits\rho D(w)\,\overline{\mathop{\rm Tr}\nolimits\sigma D(w)}\frac{%
d^{2s}w}{\left( 2\pi \right) ^{s}}  \label{parc}
\end{equation}

From now on we will consider states $\rho $ \textit{\ with finite second
moments:}%
\begin{equation*}
\sum_{j=1}^{s}\left( \mathrm{Tr}\rho q_{j}^{2}+\mathrm{Tr}\rho
p_{j}^{2}\right) \equiv \mathrm{Tr}\rho RR^{t}<\infty .
\end{equation*}%
The set of these states will be denoted $\mathfrak{S}_{2}.$ For such states
the \textit{matrix of second moments} is defined as%
\begin{equation}  \label{alpha}
\alpha ^{(2)}=\mathrm{Re}\,\mathrm{Tr}R^{t}\rho R,
\end{equation}%
and the \textit{covariance matrix} as
\begin{equation*}
\alpha =\mathrm{Re}\,\mathrm{Tr}\left( R-m\right) ^{t}\rho \left( R-m\right)
=\alpha^{(2)}-m^{t}m\leq \alpha ^{(2)},
\end{equation*}%
where $m=\mathrm{Tr}\rho R$ is the row-vector of the \textit{first moments}
(the \textit{mean vector}). It is a real symmetric $2s\times 2s$-matrix
satisfying \cite{asp}
\begin{equation}
\alpha \geq \pm \frac{i}{2}\Delta ,  \label{ur}
\end{equation}%
The state is \textit{centered} if $m=0$. For centered states the covariance
matrix and the matrix of second moments coincide and are equal to (\ref%
{alpha}).

A \textit{Gaussian state} $\rho _{m,\alpha }$ is determined by its quantum
characteristic function%
\begin{equation}
\mathrm{Tr}\,\rho _{m,\alpha }W(z)=\exp \left( im^{t}z-\frac{1}{2}%
z^{t}\alpha z\right) .  \label{gs}
\end{equation}%
Here $\alpha $ is the covariance matrix and $m$ is the mean vector. For a
centered state we denote $\rho _{\alpha }=\rho _{0,\alpha }.$

For $\rho \in \mathfrak{S}_{2}$ we have $h_{M}(\rho )\leq h_{M}(\rho
_{\alpha })<+\infty ,$ where ${\alpha }$ is the matrix of the second moments
of the state $\rho$ by the maximum entropy principle. With the quadratic
Hamiltonian
\begin{equation}  \label{qh}
H=R\epsilon R^{t},
\end{equation}
where $\epsilon$ is real positive definite $2s\times 2s$-matrix, the energy
constraint reduces to \footnote{%
We denote Sp trace of $s\times s$-matrices as distinct from trace of
operators on $\mathcal{H}$.}
\begin{equation}
\mathrm{Sp}\,\alpha \,\epsilon \,\leq E.  \label{E1}
\end{equation}

We denote the set of all states $\rho $ with the fixed matrix of second
moments $\alpha $ by $\mathfrak{S}(\alpha )$ and we will study the following
$\alpha $-\textit{constrained} capacity
\begin{equation}
C(M;\alpha )=\sup_{\mathcal{E}:\bar{\rho}_{\mathcal{E}}\in \mathfrak{S}%
(\alpha )}I(\mathcal{E},M)=\sup_{\rho \in \mathfrak{S}(\alpha )}\left[
h_{M}(\rho )-e_{M}(\rho )\right] .  \label{ca}
\end{equation}%
The energy-constrained classical capacity (\ref{cmfe}) of the measurement
channel $M$ is
\begin{equation*}
C(M;H,E)=\sup_{\alpha :\mathrm{Sp}\,\alpha \epsilon \leq E}C(M;\alpha ).
\end{equation*}

A \textit{Gaussian} measurement channel $M$ in the sense of \cite{QSCI},
\cite{hclass} is defined via the operator-valued characteristic function of
the form%
\begin{equation}
\int \mathrm{e}^{iz^{t}w}M(dz)=\exp \left( i\,R\,Kw-\frac{1}{2}w^{t}\beta
w\right) ,  \label{cf}
\end{equation}%
where $K$ is a scaling matrix, $\beta $ is the measurement noise covariance
matrix, $\beta \geq \pm \frac{i}{2}K^{t}\Delta K$. The case $K=I_{2s}$ (as
well as of a general nondegenerate $K$) corresponds to the type 1 Gaussian
measurement channel (with the multimode noisy heterodyning, see e.g \cite%
{caves}, \cite{sera} as the prototype). However (\ref{cf}) includes also
type 2 and 3 measurement channels (noisy and noiseless multimode homodyning)
in which case $K$ is a projection onto an isotropic subspace of $Z$ (i.e.
one on which the symplectic form $\Delta $ vanish). The following theorem
was proved in \cite{entropy}:

\textbf{Theorem 1.} \label{t1} \textit{Let }$M$\textit{\ be a general
Gaussian measurement channel. The optimizing density operator $\rho $ in (%
\ref{ca}) is a (centered) Gaussian density operator $\rho _{\alpha }:$%
\begin{equation}
C(M;\alpha )=h_{M}(\rho _{\alpha })-e_{M}(\rho _{\alpha }),  \label{cma}
\end{equation}%
and hence for a quadratic Hamiltonian (\ref{qh})
\begin{equation}
C(M,H,E)=\max_{\alpha :\mathrm{Sp}\,\alpha \,\epsilon \,\leq E}C(M;\alpha
)=\max_{\alpha :\mathrm{Sp}\,\alpha \,\epsilon \,\leq E}\left[ h_{M}(\rho
_{\alpha })-e_{M}(\rho _{\alpha })\right] .  \label{cmhe}
\end{equation}%
} %\end{Theorem}

The theorem asserts that the average state of an optimal encoding for a
Gaussian measurement is Gaussian but says nothing about the detailed
structure of the ensemble. It is well known that a non-Gaussian ensemble can
have Gaussian average state (a canonical example is ensemble of the Fock
states with the geometric distribution).

\textbf{Hypothesis of Gaussian Maximizers (HGM)}: \textit{Let $M$ be an
arbitrary Gaussian measurement channel. Then there exists an optimal
ensemble for (\ref{e1}) and hence for (\ref{0}) which is Gaussian, more
precisely it consists of (properly squeezed) coherent states with the
displacement parameter having Gaussian probability distribution.}

For Gaussian measurement channels of the type 1 the minimum output
differential entropy
\begin{equation}
\inf_{\rho }h_{M}(\rho )  \label{e=min}
\end{equation}%
is attained on a pure Gaussian (i.e. squeezed) state (this follows from the
result of \cite{ghm} applied to the entropy function and the complex
structure associated with $M$). If in addition the Gaussian state $\rho
_{\alpha }$ satisfies the \textquotedblleft threshold
condition\textquotedblright\ which means that the covariance matrix $\alpha $
dominates the covariance matrix of the entropy-minimizing squeezed state,
then $\rho _{\alpha }$ can be represented as a Gaussian mixture of these
squeezed states, thus giving the optimal ensemble. This implies the validity
of the HGM and an efficient computation of the $\alpha $-constrained
capacity as%
\begin{equation*}
C(M,H,E)=h_{M}(\rho _{\alpha })-\min_{\rho }h_{M}(\rho )
\end{equation*}%
see \cite{acc-noJ}. On the other hand, this does not work when the
\textquotedblleft threshold condition\textquotedblright\ is violated, and
notably, for all Gaussian measurement channels of the type 2 (noisy
homodyning), with the generic example of the energy-constrained approximate
measurement of the position $q=\left[ q_{1},\dots ,q_{s}\right] $ subject to
Gaussian noise.
%(see \cite{hy}, where the entanglement-assisted capacity of
%such a measurement was computed).
In the section \ref{s5} we will apply the method from section \ref{s2} to
prove the HGM in this case for a single mode system. Our strategy will be
the computation of $e_{M}(\rho _{\alpha })$ with the optimality conditions
of that section and relying on the formula (\ref{cmhe}).

\textbf{Remark 1.} In the case of the oscillator-type Hamiltonian%
\begin{equation*}
H=\sum_{j,k=1}^{s}\left( q\,\epsilon _{q}q^{t}+p\,\epsilon _{p}p^{t}\right) ,
\end{equation*}%
where $q=[q_{1},\dots ,q_{s}],\,p=[p_{1},\dots ,p_{s}],$ the energy
constraint is
\begin{equation}
\mathrm{Sp}\,\epsilon _{q}\alpha _{q}+\mathrm{Sp}\,\epsilon _{p}\alpha
_{p}\leq E.  \label{ed}
\end{equation}%
Then the maximization in (\ref{cmhe}) can be taken over only block-diagonal
covariance matrices $\alpha =\left[
\begin{array}{cc}
\alpha _{q} & 0 \\
0 & \alpha _{p}%
\end{array}%
\right] .$ The argument relying upon concavity of the capacity as the
function of the average state of the ensemble \cite{Shir1}, is similar to
one given in sec. 4 of \cite{hy} for the case of entanglement-assisted
capacity.

\section{The classical capacity of approximate position measurement}

\label{s4}

The approximate (unsharp) measurement of position $q$ in one mode $q,p$ (a
mathematical model for noisy homodyning) is given by POVM $M(dy)=m(y)dy$,
where
\begin{equation}
m(y)=\frac{1}{\sqrt{2\pi \beta }}\exp \left[ -\frac{\left( q-y\right) ^{2}}{%
2\beta }\right] =\frac{1}{\sqrt{2\pi \beta }}D(y)\exp \left[ -\frac{q^{2}}{%
2\beta }\right] D(y)^{\ast },  \label{apprq}
\end{equation}%
where $\beta >0$ is the power of the Gaussian noise, $D(y)=\exp \left(
-iyp\right) $. The Gaussian measurement channel given by this POVM acts on a
centered Gaussian state $\rho _{\alpha }$ with the covariance matrix $\alpha
=\left[
\begin{array}{cc}
\alpha _{q} & 0 \\
0 & \alpha _{p}%
\end{array}%
\right] $ by the formula
\begin{equation}
M:\rho _{\alpha }\rightarrow \exp \left[ -\frac{y^{2}}{2\left( \alpha
_{q}+\beta \right) }\right] \frac{dy}{\sqrt{2\pi \left( \alpha _{q}+\beta
\right) }}.  \label{1}
\end{equation}%
Take the oscillator Hamiltonian $H=\frac{1}{2}\left( q^{2}+p^{2}\right) .$
The problem is to compute the classical capacity%
\begin{equation}
C(M,H,E)=\max_{\mathcal{E}:\mathrm{Tr}\bar{\rho}_{\mathcal{E}}H\leq E}I(%
\mathcal{E},M),  \label{3}
\end{equation}%
and the maximization is over the input ensembles of states $\mathcal{E}$
(encodings) satisfying the energy constraint $\mathrm{Tr}\bar{\rho}_{%
\mathcal{E}}H\leq E.$ Remark 1 above shows that we can restrict to ensembles
with average state having the diagonal matrix $\alpha$ as above.

In other words, one makes the \textquotedblleft classical\textquotedblright\
measurement of the observable%
\begin{equation*}
Y=q+\xi ,\quad \xi \sim \mathcal{N}(0,\beta ),
\end{equation*}%
with the quantum energy constraint $\mathrm{Tr}\,\rho (q^{2}+p^{2})\leq 2E$,
aiming to transmit the maximum information. The difficulty is that one
measures $q$, while imposing the constraint on the energy $H=\frac{1}{2}%
\left( q^{2}+p^{2}\right) ,$ involving an implicit constraint on $p$ which
does not commute with $q$.

As we have mentioned, there is no general \textquotedblleft Gaussian
maximizer\textquotedblright\ result for $C$ in such cases, therefore we will
first find the maximum over (special) \textit{Gaussian ensembles}. The final
goal will be to prove the HGM showing thus that the found Gaussian ensemble
is in fact a solution of the capacity problem (\ref{3}) by checking the
optimality conditions from sec. \ref{s2}.

\textbf{HGM for approximate position measurement }\cite{nano}: the maximum
is attained on the Gaussian ensemble $\mathcal{E}_{gauss}=\left\{ \pi
_{0}(dx),\rho _{0} (x)\right\} $ where $\rho _{0} (x)=\left\vert
x\right\rangle _{\delta }\left\langle x\right\vert $ is the pure Gaussian
(squeezed) state with the vector $\left\vert x\right\rangle _{\delta
}=D(x)\left\vert 0\right\rangle _{\delta },$ and the squeezed vacuum has
zero mean and the following second moments:
\begin{equation*}
_{\delta }\left\langle 0\right\vert q^{2}\left\vert 0\right\rangle _{\delta
}=\delta ,\quad \mathrm{Re}~ _{\delta }\left\langle 0\right\vert q
p\left\vert 0\right\rangle _{\delta }=0,\quad _{\delta }\left\langle
0\right\vert p^{2}\left\vert 0\right\rangle _{\delta }=\frac{1}{4\delta }.
\end{equation*}%
The distribution $\pi _{0}(dx)=\frac{1}{\sqrt{2\pi \gamma }}\exp \left[ -%
\frac{x^{2}}{2\gamma }\right] dx.$ For the fixed centered Gaussian state $%
\rho _{\alpha }$ with the covariance matrix $\alpha =\left[
\begin{array}{cc}
\alpha _{q} & 0 \\
0 & \alpha _{p}%
\end{array}%
\right] ,$ in order that the average state of the ensemble to be $\int \rho
(x)\pi _{0}(dx)=\rho _{\alpha }$ the parameters should satisfy $\frac{1}{%
4\delta }=\alpha _{p},\,\delta +\gamma =\alpha _{q},$ whence
\begin{equation}  \label{param}
\delta =\frac{1}{4\alpha _{p}},\quad\gamma =\alpha _{q}-\frac{1}{4\alpha _{p}%
}.
\end{equation}
This ensemble encodes the information solely into the displacement $x$ of
the position leaving the momentum intact.

Similarly to (\ref{1})%
\begin{equation}  \label{pdel}
p_{\rho _{0}(x)}(y)=\frac{1}{\sqrt{2\pi \left( \beta +\delta \right) }}\exp %
\left[ -\frac{\left( y-x\right) ^{2}}{2\left( \beta +\delta \right) }\right].
\end{equation}%
Using this and (\ref{1}) we get the ``Gaussian'' values%
\begin{equation}
h_{M}(\rho _{\alpha })=\frac{1}{2}\ln \left( \alpha _{q}+\beta \right) +
\frac{1}{2}\ln (2\pi e),  \label{hm1}
\end{equation}%
\begin{equation}
e_{M}(\rho _{\alpha })=\frac{1}{2}\ln \left( \frac{1}{4\alpha _{p}}+\beta
\right) +\frac{1}{2}\ln (2\pi e),  \label{em1}
\end{equation}%
hence taking into account (\ref{cmhe}),
\begin{equation}
C_{gauss}(M;\alpha )=h_{M}(\rho _{\alpha })-e_{M}(\rho _{\alpha })=\frac{1}{2%
}\ln \frac{\alpha _{q}+\beta }{\frac{1}{4\alpha _{p}}+\beta }.  \label{cg2}
\end{equation}%
The Gaussian constrained capacity is thus
\begin{eqnarray}
C_{gauss}(M,H,E) &=&\max_{\alpha _{q}+\alpha _{q}\leq 2E}\frac{1}{2}\left[
\ln \left( \alpha _{q}+\beta \right) -\ln \left( 1/\left( 4\alpha
_{p}\right) +\beta \right) \right]  \label{cg1} \\
&=&\max_{\alpha _{p}}\frac{1}{2}\left[ \ln \left( 2E-\alpha _{p}+\beta
\right) -\ln \left( 1/\left( 4\alpha _{p}\right) +\beta \right) \right] ,
\notag
\end{eqnarray}%
where in the second line we took the maximal value $\alpha _{q}=2E-\alpha
_{p}$. Differentiating, we obtain the equation for the optimal value $\alpha
_{p}$:
\begin{equation*}
4\beta \alpha _{p}^{2}+2\alpha _{p}-\left( 2E+\beta \right) =0,
\end{equation*}%
the positive solution of which is%
\begin{equation}  \label{ap}
\alpha _{p}=\frac{1}{4\beta }\left( \sqrt{1+8E\beta +4\beta ^{2}}-1\right) ,
\end{equation}%
whence%
\begin{equation}  \label{cgaus}
C_{gauss}(M,H,E)=\ln \left( \frac{\sqrt{1+8E\beta +4\beta ^{2}}-1}{2\beta }%
\right) .
\end{equation}%
The parameters of the optimal Gaussian ensemble are obtained by substituting
the value (\ref{ap}) into (\ref{param}) with $\alpha _{q}=2E-\alpha _{p}$ ~%
\footnote{%
Notably, the expression (\ref{cgaus}) is of the same type as the one
obtained in \cite{hall3} by optimizing the information from applying sharp
position measurement to noisy optimally squeezed states (the author is
indebted to M. J. W. Hall for this observation).}.

The case of sharp position measurement $(\beta =0)$ formally corresponding
to $M(dy)=\delta (q-y)dy,$ is not included in the discussion above. Yet for $%
\beta =0$ the formula (\ref{cgaus}) gives $\delta =1/4E$ and%
\begin{equation*}
C(M,H,E)=C_{gaus}(M,H,E)=\ln 2E.
\end{equation*}%
The last formula was obtained in the paper \cite{hall2} \ where also a
general upper bound
\begin{equation}
\ln \left( 1+\frac{E-1/2}{\beta +1/2}\right) =\ln \left( \frac{2(E+\beta )}{%
1+2\beta }\right)  \label{upg}
\end{equation}%
for (\ref{3}) was given (Eq. (28) in \cite{hall2}, see also Eq. (5.39) in
\cite{caves}) .

\begin{figure}[t]
\center{\includegraphics{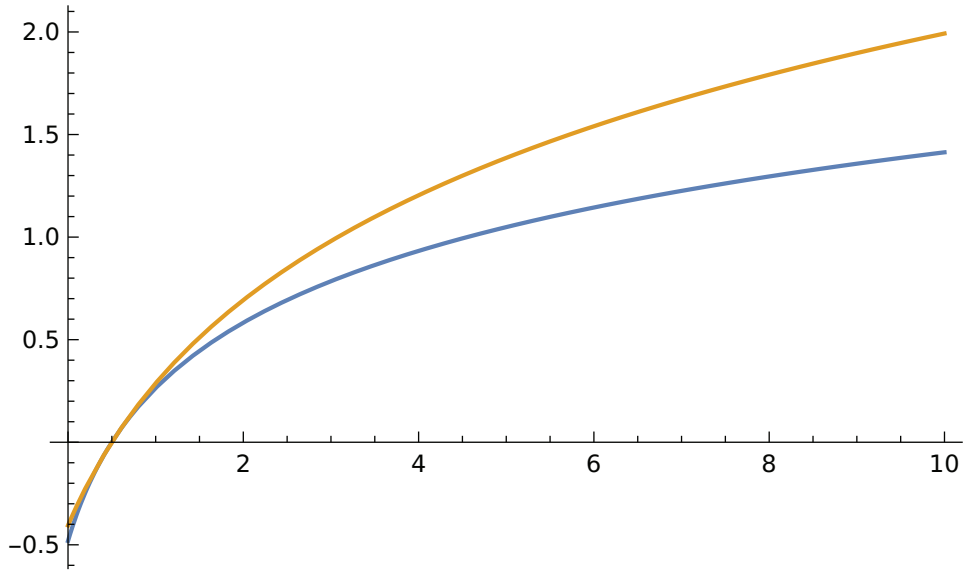}}
\caption{(color online) The Gaussian classical capacity (\protect\ref{cgaus}%
) and the upper bound (\protect\ref{upg}) ($\protect\beta =1$).}
\label{fig1}
\end{figure}

\section{Checking the optimality condition}

\label{s5}

Starting from this section it will be convenient to use natural logarithms;
then one can return to the binary logarithms if necessary.

\textbf{Theorem 2.}\label{t2} \textit{The Gaussian encoding $\mathcal{E}%
_{gauss}$ is optimal for the approximate Gaussian position measurement $M$
and the oscillator energy constraint as desribed in previous section. Its
constrained capacity $C(M,H,E)$ is equal to (\ref{cgaus}).}

\textit{Proof.} We use the method of sec. \ref{s2} and check the optimality
conditions (i), (ii) for the Gaussian encoding $\mathcal{E}_{gauss}=\left\{
\pi _{0}(dx),\rho _{0} (x)\right\} .$ We start with computation of $\Lambda
_{0}$ for $\mathcal{E}_{gauss}$.

By (\ref{K}), (\ref{apprq}), (\ref{pdel}),
\begin{eqnarray*}
K(\rho _{0}(x)) &=&\int \frac{1}{\sqrt{2\pi \beta }}\exp \left[ -\frac{%
\left( q-y\right) ^{2}}{2\beta }\right] \left[ \ln \sqrt{2\pi \left( \beta
+\delta \right) }+\frac{\left( y-x\right) ^{2}}{2\left( \beta +\delta
\right) }\right] dy \\
&=&c+\frac{\left( q-x\right) ^{2}+\beta }{2\left( \beta +\delta \right) }%
,\quad c=\ln \sqrt{2\pi \left( \beta +\delta \right) }.
\end{eqnarray*}%
Hence%
\begin{eqnarray*}
K(\rho _{0}(x))\rho _{0}(x) &=&\left[ c+\frac{\left( q-x\right) ^{2}+\beta }{%
2\left( \beta +\delta \right) }\right] D(x)\left\vert 0\right\rangle
_{\delta }\left\langle x\right\vert \\
&=&D(x)\left[ c+\frac{q^{2}+\beta }{2\left( \beta +\delta \right) }\right]
\left\vert 0\right\rangle _{\delta }\left\langle x\right\vert \\
&=&D(x)\left[ c+\frac{\beta }{2\left( \beta +\delta \right) }+\frac{\delta }{%
\left( \beta +\delta \right) }\left( \frac{q^{2}}{2\delta }+2\delta
p^{2}\right) -\frac{2\delta ^{2}p^{2}}{\left( \beta +\delta \right) }\right]
\left\vert 0\right\rangle _{\delta }\left\langle x\right\vert .
\end{eqnarray*}%
Taking into account that the squeezed vacuum $\left\vert 0\right\rangle
_{\delta }$ is the ground state for the corresponding oscillator Hamiltonian
\begin{equation*}
\left( \frac{q^{2}}{2\delta }+2\delta p^{2}\right) \left\vert 0\right\rangle
_{\delta }=\left\vert 0\right\rangle _{\delta },
\end{equation*}%
and that $D(x)$ commute with $p^{2},$ we have
\begin{equation*}
K(\rho _{0}(x))\rho _{0}(x)=\left[ c+\frac{\beta +2\delta }{2\left( \beta
+\delta \right) }-\frac{2\delta ^{2}p^{2}}{\left( \beta +\delta \right) }%
\right] \left\vert x\right\rangle _{\delta }\left\langle x\right\vert .
\end{equation*}%
Integrating over $\pi _{0}(dx),$ and taking into account that $\int
\left\vert x\right\rangle _{\delta }\left\langle x\right\vert \pi
_{0}(dx)=\rho _{\alpha },$ we obtain%
\begin{equation*}
\int K(\rho _{0}(x))\rho _{0}(x)\pi _{0}(dx)=\left[ c+\frac{\beta +2\delta }{%
2\left( \beta +\delta \right) }-\frac{2\delta ^{2}p^{2}}{\left( \beta
+\delta \right) }\right] \rho _{\alpha }.
\end{equation*}%
Comparing with (\ref{KL}), we obtain
\begin{equation}
\Lambda _{0}=c+\frac{\beta +2\delta }{2\left( \beta +\delta \right) }-\frac{%
2\delta ^{2}p^{2}}{\left( \beta +\delta \right) }.  \label{lambda0}
\end{equation}%
By condtruction, this is Hermitian operator satisfying $\left[ K(\rho
_{0}(x))-\Lambda _{0}\right] \rho _{0}(x)=0,$ i.e. the condition (ii) of
section \ref{s2}.

To check the condition (i) it is sufficient to prove
\begin{equation}
\left\langle \psi \right\vert \Lambda _{0}\left\vert \psi \right\rangle \leq
\left\langle \psi \right\vert K(\rho )\left\vert \psi \right\rangle
\label{1b}
\end{equation}%
for arbitrary density operator $\rho $ and a dense subset of $\psi \in
\mathcal{H}$. We can assume that $\psi $ is a unit vector. Since
\begin{eqnarray*}
\left\langle \psi \right\vert K(\rho )\left\vert \psi \right\rangle
&=&-\int \left\langle \psi \right\vert m(y)\left\vert \psi \right\rangle \ln
\mathrm{Tr}\rho m(y) \\
&=&-\int \left\langle \psi \right\vert m(y)\left\vert \psi \right\rangle \ln
\left\langle \psi \right\vert m(y)\left\vert \psi \right\rangle  \\
&+&\int \left\langle \psi \right\vert m(y)\left\vert \psi \right\rangle \ln
\frac{\left\langle \psi \right\vert m(y)\left\vert \psi \right\rangle }{%
\mathrm{Tr}\rho m(y)} \\
&\geq &-\int \left\langle \psi \right\vert m(y)\left\vert \psi \right\rangle
\ln \left\langle \psi \right\vert m(y)\left\vert \psi \right\rangle ,
\end{eqnarray*}%
due to nonnegativity of the relative entropy of the two probability
densities. Thus (\ref{1b}) will follow if we prove
\begin{equation}
\left\langle \psi \right\vert \Lambda _{0}\left\vert \psi \right\rangle \leq
-\int \left\langle \psi \right\vert m(y)\left\vert \psi \right\rangle \ln
\left\langle \psi \right\vert m(y)\left\vert \psi \right\rangle dy
\label{!c}
\end{equation}%
for unit vectors $\psi .$ With $\Lambda _{0}$ given by (\ref{lambda0}) it
amounts to
\begin{equation*}
\int \left\langle \psi \right\vert m(y)\left\vert \psi \right\rangle \ln
\left\langle \psi \right\vert m(y)\left\vert \psi \right\rangle dy+\ln \sqrt{%
2\pi \left( \beta +\delta \right) }+\frac{\beta +2\delta }{2\left( \beta
+\delta \right) }
\end{equation*}%
\begin{equation}
\leq \frac{2\delta ^{2}}{\left( \beta +\delta \right) }\int \left\vert \psi
\prime (x)\right\vert ^{2}dx.  \label{logsob}
\end{equation}%
The proof of this inequality is the subject of the following section.

\section{A generalization of log-Sobolev inequality}

\label{s6}

\bigskip Let $f(x)=\left\vert \psi (x)\right\vert ^{2}$ be a smooth
probability density on $\mathbb{R}$ and
\begin{equation*}
T_{t}f(y)=\frac{1}{\sqrt{2\pi t}}\int \exp \left( -\frac{(y-x)^{2}}{2t}%
\right) f(x)dx.
\end{equation*}%
Then the inequality we wish to prove, replacing $\beta $ by $t$:
\begin{equation}
\int T_{t}f(y)\ln T_{t}f(y)dy+\ln \sqrt{2\pi e\left( t+\delta \right) }+%
\frac{\delta }{2\left( t+\delta \right) }\leq \frac{2\delta ^{2}}{\left(
t+\delta \right) }\int \left\vert \psi \prime (x)\right\vert ^{2}dx
\label{conj}
\end{equation}%
for $t,\delta \geq 0.$ For $t=0,\delta =1$ this is the logarithmic Sobolev
inequality \cite{ledoux}. For $t=0,\delta >0$ it can be obtained by a change
of variable (see also (\ref{ll1}) below).

\textit{Proof} of (\ref{conj}). We start from the version of the log-Sobolev
inequality in \cite{lieb} (with dimensionality $n=1$):%
\begin{equation}
\int \left\vert \psi (x)\right\vert ^{2}\ln \frac{\,\left\vert \psi
(x)\right\vert ^{2}}{\left\Vert \psi \right\Vert _{2}^{2}}\,dx+\ln a+1\leq
\frac{a^{2}}{\pi }\int \left\vert \psi \prime (x)\right\vert ^{2}dx.
\label{ll}
\end{equation}%
Let $\left\Vert \psi \right\Vert _{2}=1$ then $f(x)=\left\vert \psi
(x)\right\vert ^{2}$ is a probability density, $\int f(x)dx=1.$ Also take $a=%
\sqrt{2\pi \delta },$ then (\ref{ll}) becomes%
\begin{equation}
\int f(x)\ln f(x)dx+\ln \sqrt{2\pi \delta }+1\leq 2\delta \int \left\vert
\psi \prime (x)\right\vert ^{2}dx  \label{ll1}
\end{equation}%
which is the same as (\ref{conj}) for $t=0.$

Denote%
\begin{equation*}
F(t,\delta )=\left( t+\delta \right) \int T_{t}f(x)\ln T_{t}f(x)dx
\end{equation*}%
\begin{equation*}
+\left( t+\delta \right) \ln \sqrt{2\pi e\left( t+\delta \right) }+\frac{%
\delta }{2}-2\delta ^{2}\int \left\vert \psi \prime (x)\right\vert ^{2}dx.
\end{equation*}%
We have to prove
\begin{equation}
F(t,\delta )\leq 0;\quad t,\delta >0.  \label{F}
\end{equation}%
We have just proved that $F(0,\delta )\leq 0.$ If we prove that $\frac{%
\partial }{\partial t}F(t,\delta )\leq 0,$ then (\ref{F}) and hence (\ref%
{conj}) will follow. We have%
\begin{equation*}
\frac{\partial }{\partial t}F(t,\delta )=\int T_{t}f(x)\ln T_{t}f(x)dx
\end{equation*}%
\begin{equation*}
+\left( t+\delta \right) \int \left[ \ln T_{t}f(x)+1\right] \frac{\partial }{%
\partial t}T_{t}f(x)\,dx+\ln \sqrt{2\pi e\left( t+\delta \right) }+\frac{1}{2%
}.
\end{equation*}%
Taking into account that $\frac{\partial }{\partial t}T_{t}f(x)=\frac{1}{2}%
\frac{\partial ^{2}}{\partial x^{2}}T_{t}f(x)$ and integrating by parts in
the second integral, we can transform it as%
\begin{eqnarray*}
&&\int \left[ \ln T_{t}f(x)+1\right] \frac{\partial }{\partial t}%
T_{t}f(x)\,dx \\
&=&-\frac{1}{2}\int \frac{\partial }{\partial x}\left[ \ln T_{t}f(x)+1\right]
\frac{\partial }{\partial x}T_{t}f(x)\,dx \\
&=&-2\int \left\vert \frac{\partial }{\partial x}\sqrt{T_{t}f(x)}\right\vert
^{2}\,dx.
\end{eqnarray*}%
Denote $g(x)=T_{t}f(x),$ then it is also a probability density, and denoting
$t+\delta =\tilde{\delta}$ we obtain%
\begin{equation*}
\frac{\partial }{\partial t}F(t,\delta )=\int g(x)\ln g(x)dx+\ln \sqrt{2\pi
\tilde{\delta}}+1-2\tilde{\delta}\int \left\vert \frac{d}{dx}\sqrt{g(x)}%
\right\vert ^{2}dx.
\end{equation*}%
Hovewer by (\ref{ll1}) this is nonpositive. Thus (\ref{F}) and hence (\ref%
{conj}) follows. This also completes the proof of theorem 2.

\section{Comment on the proof of the estimate for the convex closure of the
output entropy}

\bigskip The sufficient conditions for optimality from section \ref{s2} were
applied in our proof of theorem \ref{t2} to unbounded operators and thus
require a corresponding refinement. While this can be done in general, here
we wish point out that given the inequality (\ref{logsob}), there is
another, direct way to rigorous proof of  theorem \ref{t2}. Then the merit
of the convex programming approach is in that it allowed to generate the
conjectured inequality (\ref{logsob}).

First, we note that  (\ref{logsob}) can be extended to functions $\psi $
from  the Sobolev space  $H^{1}(\mathbb{R})$, which are square-integrable
along with its first generalized derivative $\psi ^{\prime }.$ This space is
the natural domain of definition of the momentum operator $p.$

To complete the proof of theorem \ref{t2}, in view of (\ref{cma}) and (\ref%
{hm1}), we have only to prove that%
\begin{equation}
e_{M}(\rho _{\alpha })\equiv \inf_{\mathcal{E}:\bar{\rho}_{\mathcal{E}}=\rho
_{\alpha }}\int h_{M}(\rho (\xi ))\pi (d\xi )=\frac{1}{2}\ln \left( \frac{1}{%
4\alpha _{p}}+\beta \right) +\frac{1}{2}\ln 2\pi e,  \label{ei}
\end{equation}%
where the infimum is taken over encodings $\mathcal{E=}\left\{ \pi (d\xi
),\rho (\xi )\right\} $ satisfying $\bar{\rho}_{\mathcal{E}}=\rho _{\alpha }.
$ The concavity of $h_{M}(\rho )$ implies that we can restrict to ensembles
of pure states $\rho (\xi )=\left\vert \psi _{\xi }\right\rangle
\left\langle \psi _{\xi }\right\vert ,$ so that
\begin{equation*}
\int \left\vert \psi _{\xi }\right\rangle \left\langle \psi _{\xi
}\right\vert \,\pi (d\xi )=\rho _{\alpha },
\end{equation*}%
since we can always perform the convex decomposition for all density
operators $\rho (\xi )$ into pure states without changing the barycenter and
without increasing the value of the minimized functional. It follows that
\begin{equation}
\int \left\Vert p\psi _{\xi }\right\Vert ^{2}\,\pi (d\xi )=\mathrm{Tr\,}\rho
_{\alpha }p^{2}<\infty ,  \label{ave}
\end{equation}%
hence $\psi _{\xi }\in H^{1}(\mathbb{R})$ \ for $\pi -$almost all $\xi .$
Applying the inequality (\ref{logsob}) to $\psi _{\xi }$ and rearranging
terms, we get%
\begin{equation}
h_{M}(\left\vert \psi _{\xi }\right\rangle \left\langle \psi _{\xi
}\right\vert \,)\geq \ln \sqrt{2\pi \left( \beta +\delta \right) }+\frac{%
\beta +2\delta }{2\left( \beta +\delta \right) }-\frac{2\delta ^{2}}{\beta
+\delta }\left\Vert p\psi _{\xi }\right\Vert ^{2}\,.  \label{emx}
\end{equation}%
Integrating with respect to $\pi (d\xi )$ and taking into account (\ref{ave}%
) we get
\begin{equation*}
\int h_{M}(\rho (\xi ))\pi (d\xi )\geq \ln \sqrt{2\pi \left( \beta +\delta
\right) }+\frac{\beta +2\delta }{2\left( \beta +\delta \right) }-\frac{%
2\delta ^{2}}{\beta +\delta }\alpha _{p}.
\end{equation*}%
With $\delta =\frac{1}{4\alpha _{p}}$ we get the value at the right-hanf
side of (\ref{ei}), which is attained for the encoding of theorem \ref{t2}.
This proves (\ref{ei}) and hence the theorem.

Similar comment applies to the proof of HGM for approximate joint position-momentum
measurement channel (noisy heterodyning) in our subsequent e-print arXiv:2206.02133.

\section*{Appendix}

Let us illustrate the inequality (\ref{conj}) for Gaussians
\begin{equation}
f(x)=\frac{1}{\sqrt{2\pi a}}\int \exp \left( -\frac{x^{2}}{2a}\right) .
\label{gaus}
\end{equation}%
Then (\ref{conj}) reduces to%
\begin{equation*}
\ln \left( \frac{t+a}{t+\delta }\right) \geq \frac{\delta }{t+\delta }\left(
1-\frac{\delta }{a}\right)
\end{equation*}%
or introducing $u=a/t,\,v=\delta /t,$%
\begin{equation}  \label{ineq1}
\ln \left( \frac{1+u}{1+v}\right) -\frac{v}{1+v}\left( 1-\frac{v}{u}\right)
\geq 0;\quad u,v\geq 0.
\end{equation}%
To prove this inequality, notice it becomes equality for $u=\,v.$ The
derivative $d/du$ is%
\begin{equation}
\frac{1}{1+u}-\frac{v^{2}}{\left( 1+v\right) u^{2}}=\frac{\left( u-v\right)
\left( u+v+uv\right) }{\left( 1+u\right) \left( 1+v\right) u^{2}}
\label{ineq}
\end{equation}%
which is $\geq 0\,\,(\leq 0)$ if $u\geq v\,\left( u\leq v\right) .$ Hence (%
\ref{ineq1}) follows. Also we have obtained that (\ref{conj}) is exact: it
turns into equality for Gaussian (\ref{gaus}) with $a=\delta $.

%\textit{Acknowledgment.} The author is grateful to Dr. Ludovoco Lami for
%comments fostering improvement of presentation.

\end{document}